\documentclass[journal,comsoc]{IEEEtran}

\usepackage[utf8]{inputenc}
\usepackage[T1]{fontenc}
\usepackage{amsmath,amssymb} \interdisplaylinepenalty=2500
\usepackage[cmintegrals]{newtxmath} 
\usepackage{cite}
\usepackage{float}
\usepackage{graphicx} 
\usepackage{hyperref}
\usepackage{algorithm}
\usepackage{algorithmic}
\usepackage[labelformat=simple]{subcaption}

\newtheorem{example}{Example}

\ifdefined\IEEEproof

  \def\qed{\endIEEEproof}
\else
  \newenvironment{proof}{\textit{Proof: }}{\qed}
  \def\qed{\null\nobreak\hfill\ensuremath{\blacksquare}}
\fi

\makeatletter
\renewcommand*\env@matrix[1][*\c@MaxMatrixCols c]{%
	\hskip -\arraycolsep
	\let\@ifnextchar\new@ifnextchar
	\array{#1}}
\makeatother

\newcommand{\RR}{\mathbb{R}}

\newcommand{\bzero}{\mathbf{0}}
\newcommand{\bone}{\mathbf{1}}

\newcommand{\bc}{\mathbf{c}}

\newcommand{\be}{\mathbf{e}}

\newcommand{\bs}{\mathbf{s}}

\newcommand{\bx}{\mathbf{x}}
\newcommand{\by}{\mathbf{y}}
\newcommand{\bz}{\mathbf{z}}

\newcommand{\bH}{\mathbf{H}}
\newcommand{\bI}{\mathbf{I}}

\newcommand{\calC}{\mathcal{C}}

\newcommand{\calE}{\mathcal{E}}

\newcommand{\calN}{\mathcal{N}}

\usepackage{xcolor}

\begin{document}

\title{Iterative Error Decimation for Syndrome-Based Neural Network Decoders}

\author{Jorge Kysnney Santos~Kamassury and Danilo~Silva

\thanks{J. K. S. Kamassury and D. Silva are with the Department of Electrical and Electronic Engineering, Federal University of Santa Catarina, Florianópolis-SC, Brazil (e-mail: jorge.kamassury@posgrad.ufsc.br; danilo.silva@ufsc.br). 
	
This work was partially supported by CNPq under Grant 132881/2018-7.}
}

\maketitle 
\begin{abstract}
In this letter, we introduce a new syndrome-based decoder where a deep neural network (DNN) estimates the error pattern from the reliability and syndrome of the received vector. The proposed algorithm works by iteratively selecting the most confident positions to be the error bits of the error pattern, updating the vector received when a new position of the error pattern is selected. Simulation results for the (63,45) and (63,36) BCH codes show that the proposed approach outperforms existing neural network decoders. In addition, the new decoder is flexible in that it can be applied on top of any existing syndrome-based DNN decoder without retraining.
\end{abstract}

\section{Introduction}
\IEEEPARstart{I}{n} recent years, investigations into the design of short-length channel codes have acquired notability, particularly due to applications that newer technologies aim to support. 5G technology, in particular, aims to guarantee services that require ultra-reliable low-latency communication (URLLC) \cite{huang}. For example, intelligent transport systems and process automation demand reliability in the order of $10^{-3} $ to $10^{-6}$ and latency between 1 ms to 100 ms. Communication under these conditions is challenging, since the requirements themselves are strict and conflicting \cite{shirv,durisi}.

This scenario has motivated the evaluation possible candidate codes in terms of reliability and complexity for a given (short) blocklength \cite{liva_code, shirv,efficient}. Among many candidates---which include polar codes, LDPC codes and convolutional codes---BCH codes stand out as having an excellent performance, very close to the fundamental limits in the short blocklength regime. This is achieved by the use of an ordered statistics decoder (OSD), which delivers near-maximum-likelihood (ML) performance; however, this comes at the price of a high complexity, which grows quickly as the blocklength increases.

An alternative that has increasingly been explored in recent work is the use of decoders based on deep neural networks (DNNs). Although the use of neural networks (NNs) for the task of decoding is not recent \cite{tallini}, due to the success of deep learning in several applications, interest in this purpose has been resumed \cite{gruber}. Recently, in \cite{nachmani_learning}, Nachmani \textit {et al.} proposed a deep learning framework that is modeled on the LDPC belief propagation (BP) decoder, where connections between neurons (as well as activations) are designed to mimic the underlying Tanner graph. In subsequent works \cite{nachmani, off_set, lugosch, beery}, other architectures based on \cite{nachmani_learning} are presented.

Unlike approaches based on BP decoding, Bennatan \textit{et al.} proposed in \cite{bennatan} a new decoder structure, where the NN is fed the reliability and syndromes of the received sequences and acts on noise estimation. Their approach can be regarded as a soft-decision extension of the syndrome-based approach of \cite{tallini}. 
A great advantage of this structure is that the NN can be designed freely, i.e., without the restrictions present in architectures based on the BP decoder. Subsequently, the vanilla DNN proposed in \cite{bennatan} was simplified in \cite{kavv1,kavv2}; specifically, the architecture in \cite{kavv2} has fewer parameters and achieves a better performance than the original one.

A common limitation in many previous works is their focus on the bit error rate (BER) as a measure of performance, presumably because it maps more directly to the NN training objective. However, when evaluated by the block error rate (BLER), some of these works fail to significantly improve upon a hard-decision bounded-distance decoder (HD-BDD) that would conventionally be used to decode BCH codes.

In this paper, we present a strategy to improve the performance of any syndrome-based neural decoder (i.e., any decoder following the approach in \cite{bennatan}), at the expense of a moderate increase in complexity. 

Our approach is to take the unquantized estimate of the error vector that is output by a neural decoder and iteratively select its most confident position, which is then \textit{decimated} (subtracted) from the received vector before a new decoding attempt is made.
Our results show that this proposed approach significantly improves the BLER achieved 
by the decoder in \cite{bennatan}, outperforming previous results for the BCH(63,36) and BCH(63,45) codes. \footnote{Code available at \url{https://github.com/Kamassury/IED}.}

\subsubsection*{Notation} 
We use $x_i$ for the $i$th element of a vector $\bx$. Let $\bzero$ and $\bone$ be the all-zeros and the all-ones vectors, respectively, with lengths implied by the context. If $\bx \in \RR^n$ and $\gamma \in \RR$, then $1[\bx > \gamma]$ denotes the vector $\by \in \{0,1\}^n$ such that $y_i = 1$ if and only if $x_i > \gamma$. We use a similar notation for $1[\bx < \gamma]$.    

\section{Preliminaries} \label{sec:preliminaries}
\subsection{Channel model} \label{sec:channel}

Let $\calC \subseteq \{0,1\}^n$ be an $(n,k)$ binary linear code with parity-check matrix $\bH \in \{0,1\}^{(n-k)\times n}$. Suppose a codeword $\bc \in \calC$ chosen uniformly at random is transmitted over a binary-input additive white Gaussian noise (BI-AWGN) channel. The received vector is given by
\begin{equation}
\by = \bone - 2 \bc + \bz
\end{equation}
where $\bz \sim \calN(\bzero, \sigma^2 \bI_n)$ and $\sigma^2 = N_0/(2E_b)$. The goal of the decoder is to infer $\bc$ from $\by$, producing an estimate $\hat{\bc} \in \{0,1\}^n$. The block error probability (BLER) is defined as $P[\hat{\bc}\neq \bc]$.

\subsection{Syndrome-Based Neural Decoding}\label{sec:syndrome}
Let $\by_b = 1[\by < 0] \in \{0,1\}^n$ be the vector of hard decisions\footnote{Note that $y_i = -1 + z_i$ when $c_i=1$.} 
and let $\be = \by_b + \bc \bmod 2 \in \{0,1\}^n$ be corresponding error vector. Clearly, $\bc$ can be easily found given $\by_b$ and $\be$. Thus, the decoding problem reduces to that of estimating $\be$.
As shown in \cite{bennatan}, a sufficient statistic for the estimation of~$\be$ is the pair $(\bs, |\by|)$, where $\bs = \by_b \bH^T \bmod 2$ is the \textit{syndrome} of the error vector (i.e., $\bs = \be \bH^T \bmod 2$) and $|\by|=(|y_1|,\ldots,|y_n|)$ is the vector of channel reliabilities. 

\begin{figure}
	\centering
	\includegraphics[scale=0.69]{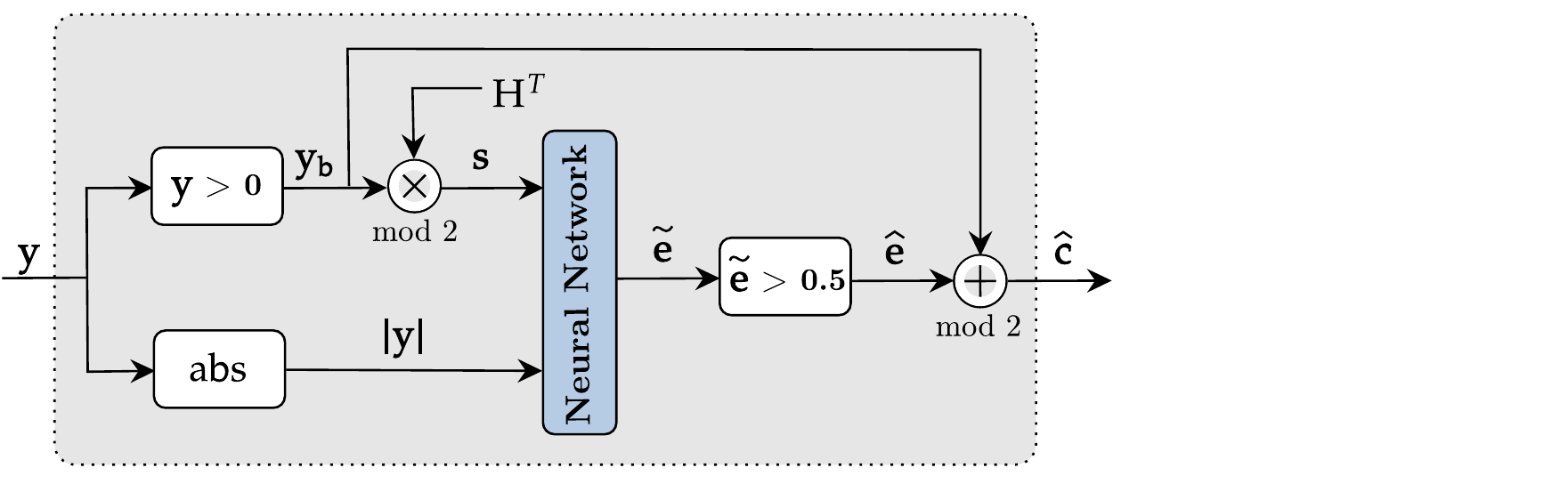}
	\caption{A general syndrome-based neural decoder.}
	\label{fig:decoder}
\end{figure}

The approach proposed in \cite{bennatan} is to design an NN to estimate~$\be$ from $(\bs, |\by|)$. More precisely, the network is trained to minimize the empirical risk $E[\sum_{i=1}^n L(e_i,\tilde{e}_i)]$ under the channel distribution, where $L(e_i, \tilde{e}_i) = -e_i \log \tilde{e}_i - (1-e_i) \log (1-\tilde{e}_i)$ is the binary cross-entropy (BCE) loss function and $\tilde{\be} \in [0,1]^n$ is the NN output, produced with a sigmoid output activation function.\footnote{The original description in \cite{bennatan} uses a $[-1,1]$ mapping and a hyperbolic tangent output activation function, which is mathematically equivalent to the description given here.} The binary estimate of~$\be$ is then obtained as $\hat{\be} = 1[\tilde{\be} > 0.5] \in \{0,1\}^n$. The complete decoder, which we refer to as a syndrome-based neural decoder (SBND), is shown in Fig.~\ref{fig:decoder}.

As argued in \cite{bennatan}, the inputs $(\bs,|\by|)$ and the target~$\be$ are all independent of $\bc$, thus the zero codeword assumption $\bc=\bzero$ can be used for both training and performance evaluation of the decoder. This avoids the risk of overfitting to the subset of codewords used during training. Moreover, as with any neural decoder, since the channel model is known, a potentially unlimited number examples can be used for training and testing without risk of overfitting to the noise.

\section{Iterative Error Decimation Decoder}\label{sec:neural-max}

\subsection{Motivation}
\label{ssec:ied-motivation}
A main issue in training a syndrome-based neural decoder according to the procedure in Section \ref{sec:syndrome} is the potential presence of inconsistent (or ``noisy'') training examples, namely, training examples with the same (or very similar) inputs but different targets. This phenomenon, called \textit{disturbance} in \cite{tallini}, is most clearly seen in a decoder where the input component $|\by|$ is removed from the neural network, i.e., the neural network is trained to predict the target error vector $\be$ solely from its syndrome $\bs$. 
Note that this corresponds to degrading the BI-AWGN channel into a binary symmetric channel (BSC), which is the channel originally considered in \cite{tallini}. In this case, multiple target error vectors with the same syndrome are likely to appear during training, producing a ``noisy'' output that tends to be a superposition of those error vectors. 

For simplicity, consider the BSC case in the following. Ideally, the neural network should be trained to emulate the performance of a maximum-likelihood decoder; thus, every syndrome $\bs$ should be paired with a \textit{single} lowest-weight error vector $\be$ corresponding to that syndrome, in order to form the training set. Any distinct error vector with the same syndrome, if used as a training example, will drive the network to deviate from the desired prediction and thus can only hurt performance. However, generating such an optimal training set requires performing maximum-likelihood decoding for every possible syndrome (or, equivalently, generating and storing a full syndrome table) which can be computationally infeasible.

A simple approach proposed in \cite{tallini} to avoid disturbance is to restrict the training set to only target error vectors of weight up to the guaranteed error-correction capability of the code, $t = \lfloor (d_{\min}-1)/2 \rfloor$, where $d_{\min}$ is the minimum distance of the code. This set is guaranteed to have a single error vector for each syndrome. However, under this approach, 
the neural network is unlikely to learn to predict error vectors of larger weights, which is precisely what is needed in order to outperform a bounded-distance decoder.

Now, let us illustrate what can happen when an inconsistent training set is used. For instance, consider a $(15, 5, 7)$ BCH code. For this code, the error vectors
\begin{align*}
\be_1 &= (1, 1, 1, 1, 0, 0, 0, 0, 0, 0, 0, 0, 0, 0, 0) \\
\be_2 &= (0, 0, 0, 0, 0, 1, 0, 1, 1, 0, 0, 1, 0, 0, 0)
\end{align*}
have exactly the same syndrome (and these are the only lowest-weight vectors with that syndrome). For that syndrome, the output of an NN 
(trained as in Section \ref{sec:syndrome}) may be, e.g.,
\begin{multline*}
\tilde{\be} = (0.479, 0.505, 0.512, 0.491, 0.005, 0.507, 0.000, 0.516,\\ 0.481, 0.000, 0.000, 0.483, 0.002, 0.001, 0.000)
\end{multline*}
which, after thresholding at $0.5$, leads to the estimate 
\begin{equation*}
\hat{\be} = (0, 1, 1, 0, 0, 1, 0, 1, 0, 0, 0, 0, 0, 0, 0).
\end{equation*}
This prediction is always incorrect, as it does not even correspond to the input syndrome.

An explanation for this behavior is that, under the architecture and training approach of Section \ref{sec:syndrome}, the NN is modeling the \textit{bitwise} posterior probability
\[
\tilde{e}_j \approx P[e_j=1 \mid (\bs, |\by|)].
\]
While this approach can potentially lead to a low bit-error rate (BER), it is clearly unsuited to obtaining low BLER. On the other hand, regarding the problem as a multiclass classification among all possible error vectors (e.g., using softmax output activation with categorical cross-entropy loss) \cite{low} is clearly computationally infeasible unless $n$ is very small.

\subsection{Iterative Error Decimation}\label{sec:ied}
Rather than modifying the training procedure to avoid disturbance as in \cite{tallini}, we propose to modify the decoder so as to make it robust to the superposition of error patterns. 

Our approach is to perform $T-1$ iterations where a \textit{single} bit is selected that is  most likely (as estimated by the neural network) to be in error; this bit is then flipped in the received vector and the decoding is repeated, until the $T$th iteration where thresholding at 0.5 is applied. We call this procedure \textit{iterative error decimation} (IED). The underlying idea is that, after a bit error is (correctly) eliminated, the resulting problem becomes easier to solve, leading to more confident estimates. 
Note that IED can be applied to any syndrome-based neural decoder, without requiring any changes in the training stage.


A detailed description of the decoder is given in Algorithm~\ref{algoritmo:2}. Note that, in line 8, we assume that the NN outputs probability estimates. In line 10, we select the position $j$ of the largest (thus, most confident) element of the vector $\tilde{\mathbf{e}}$. The decimation step occurs at line 11, where the sign of the received vector is flipped at the position~$j$ estimated to be in error. Since we assume certainty that the chosen position is in error, in principle we could also set the magnitude $|\by_j|$ to infinity (or to a very large value). However, in our experiments we observed that setting $|\by_j|$ to a too high value actually hurts performance, possibly because such values were not observed during training. In practice, we found that the best results are obtained when we do not change the magnitude of $|\by_j|$.

The algorithm stops when a zero syndrome has been obtained (line 4) or when $T$ iterations have been performed, at which point thresholding is applied to the remaining error estimate.

Clearly, the complexity of one iteration of the IED decoder is dominated by that of the NN inference step. Since the number of iterations is at most $T$, the maximum latency is at most $T$ times that of a conventional SBND. On the other hand, the average number of iterations is upper bounded by
\[
1 + P[\calE_1] + \cdots + P[\calE_1,...,\calE_{T-1}] \leq 1 + P[\calE_1] + ... + P[\calE_{T-1}]
\]
where $P[\calE_i]$ is the block error probability of an IED decoder with $i$ iterations. Thus, compared to a conventional SBND, the relative increase in the average complexity is typically very small and becomes negligible for high $E_b/N_0$.

\begin{algorithm}[ht]
	\caption{Iterative error decimation (IED) decoder}
	\label{algoritmo:2}
	\begin{algorithmic}[1]
		\REQUIRE $\mathbf{y},\,\hbox{H},\,\hbox{T}$
		\ENSURE $\hat{\mathbf{c}}$
		\FOR{$i = 1, \ldots, \hbox{T}$}
		\STATE $\mathbf{y_{b}}\leftarrow 1[\mathbf{y}<0]$
		\STATE $\mathbf{s}\leftarrow \mathbf{y_{b}}\hbox{H}^{T} \bmod 2$
		\IF{$\mathbf{s}=0$}
		\STATE $\hat{\mathbf{c}}\leftarrow \mathbf{y_{b}}$
		\RETURN $\hat{\mathbf{c}}$
		\ENDIF
		\STATE $\tilde{\mathbf{e}}\leftarrow\hbox{NN}(\mathbf{s}, |\mathbf{y}|)$
		\IF{$i<\hbox{T}$}
		\STATE $j\leftarrow \arg \max (\tilde{\mathbf{e}})$
		\STATE $\mathbf{y}_j\leftarrow -\mathbf{y}_j$
		\ENDIF
		\ENDFOR
		\STATE $\hat{\mathbf{e}}\leftarrow 1[\tilde{\mathbf{e}}>0.5]$
		\STATE $\hat{\mathbf{c}}\leftarrow \mathbf{y_{b}} + \hat{\mathbf{e}} \bmod 2$
		\RETURN $\hat{\mathbf{c}}$
	\end{algorithmic}
\end{algorithm}

\section{Experiments and results} \label{sec:results}

In this section, we investigate the BLER performance of the decoders described in the sections \ref{sec:syndrome} and \ref{sec:ied} for the linear codes BCH(63,45) and BCH(63,36), where BCH$(n,k)$ denotes a primitive narrow-sense binary BCH code of length~$n$ and dimension~$k$. For comparison purposes, we use the best results obtained in \cite {lugosch, beery, kavv2} as well as the HD-BDD 
and ML \cite{ml} performances. With respect to BER performance, we compare specifically with \cite{beery}, \cite{kavv2} and \cite{near_maximum} (note that \cite{lugosch} presents only BLER performance). All simulations were performed using the Keras API with Tensorflow backend.

For the training of DNNs, we have used $10^{7}$ examples (generated in real time) with $E_{b}/N_{0}=4$ dB. This value of $E_{b}/N_{0}$ is suggested in \cite{gruber} to give a good balance between noise and code structure in the training examples presented for the DNN to learn. We have used Glorot normal 
initialization and the Adam optimizer with batches of size 2048.

In the inference stage, the BLER was estimated by running Monte Carlo simulations until the occurrence of at least 100 block errors for each value $E_{b}/N_{0}$.

\subsection{BCH(63,45) code}
\label{ssec:experiments-45}
For the BCH(63,45) code, we use the DNN architecture presented in \cite{kavv2}, which has seven fully connected layers. The six hidden layers have 300 units each and use a rectified linear unit (ReLU) as activation function \cite{fellow}. 

Following the same procedures described in \cite {kavv2}, for this architeture the learning rate for the gradient propagation is initialized to $10^{-3}$ and is reduced by a factor of $10^{-1}$ when the validation loss stops reducing for 5 epochs. 

\begin{figure}[htbp!]
	\centering
	\begin{subfigure}[b]{0.48\textwidth}
		\centering
		\includegraphics[width=0.95\linewidth]{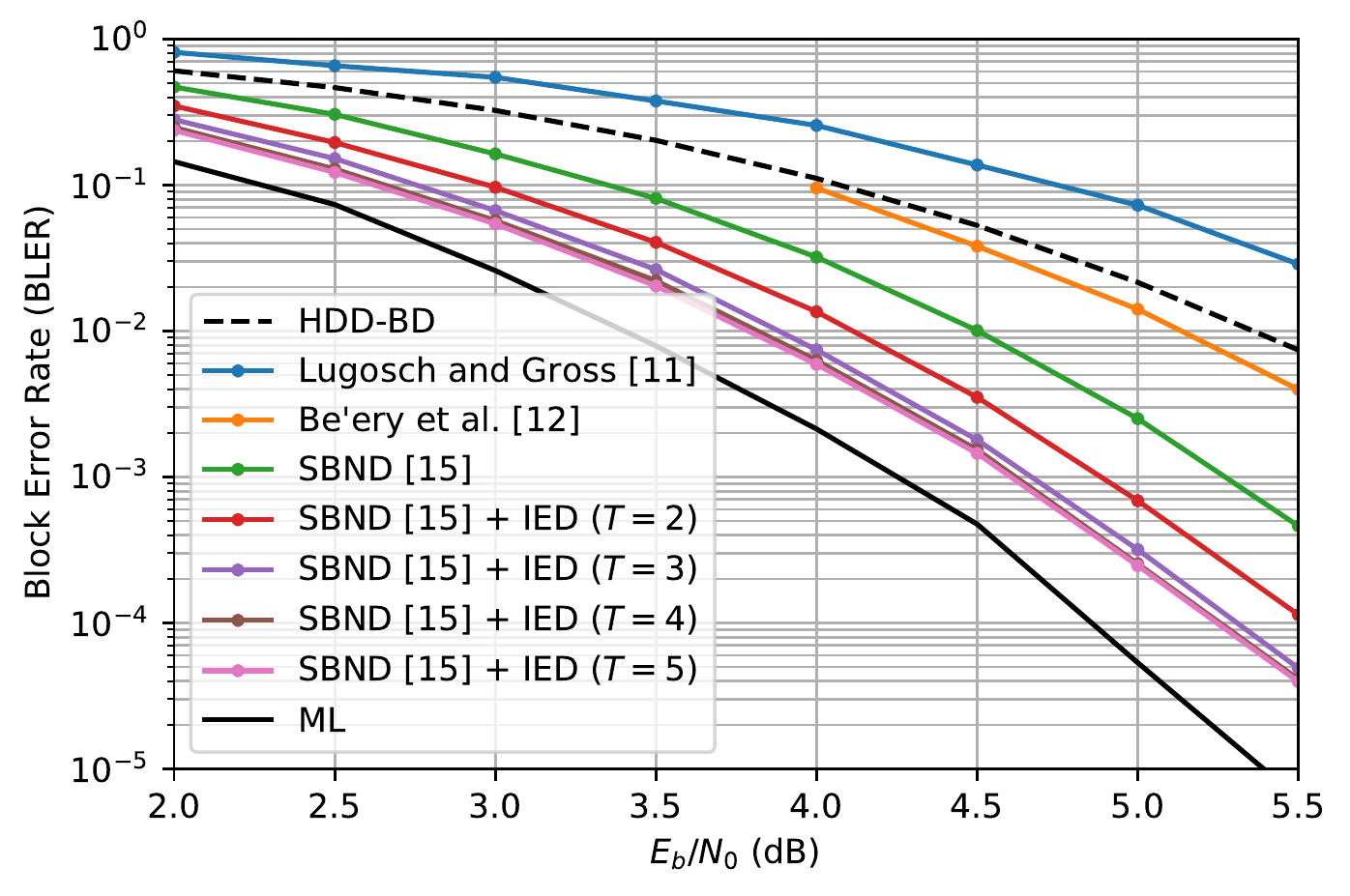}
		\centering
			\caption{}
		\label{fig:res_Ia}
	\end{subfigure}
\hfill
	\begin{subfigure}[b]{0.48\textwidth}
	\centering
	\includegraphics[width=0.95\linewidth]{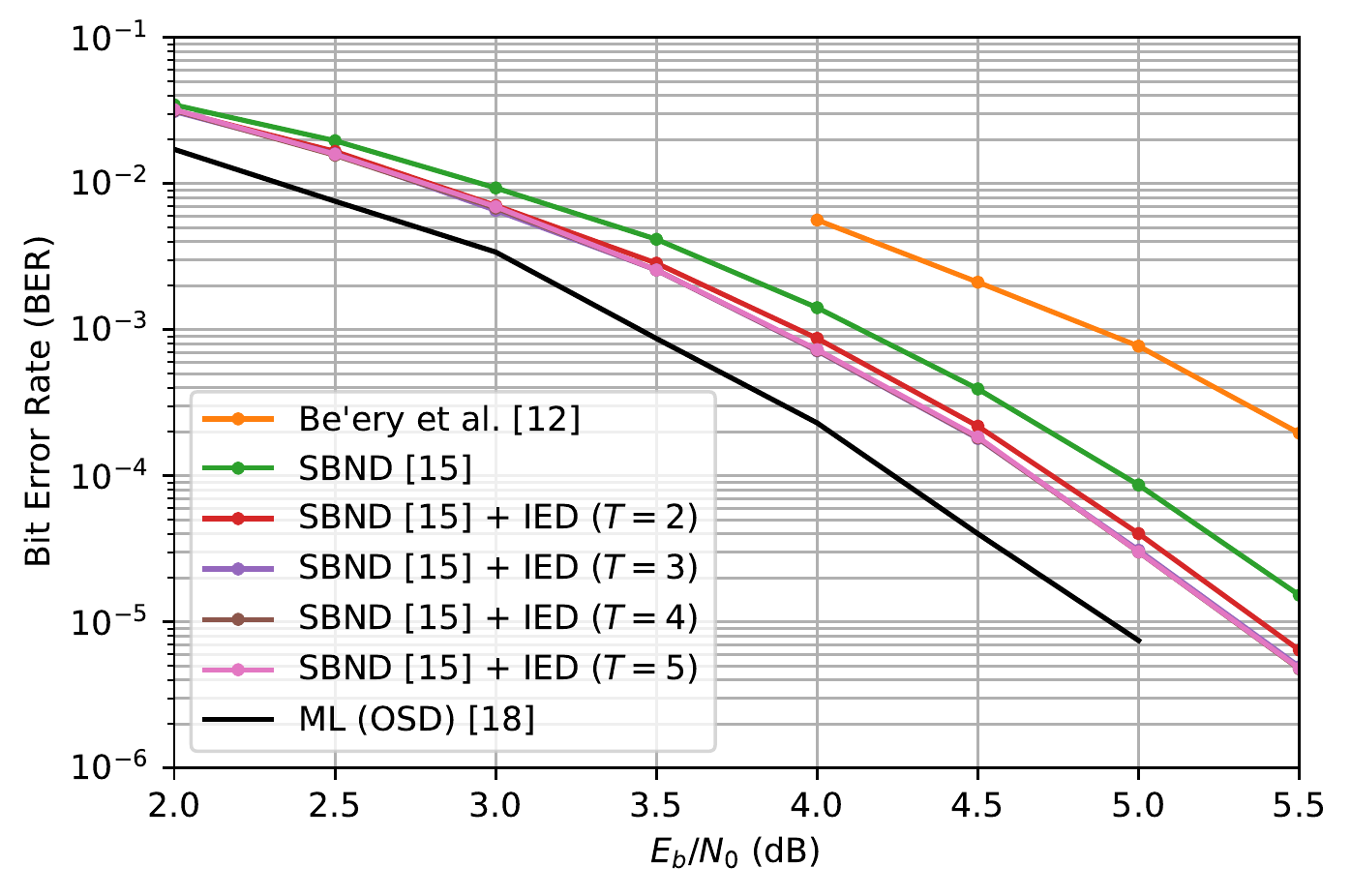}
	\centering
	\caption{}
	\label{fig:res_Ib}
\end{subfigure}
	\caption{Performance obtained with the decoder in \cite{bennatan} and the IED decoder for the BCH(63,45) code, using the DNN in \cite{kavv2}.}
\label{fig:res_I}
\end{figure}

Fig. \ref{fig:res_I} shows the performance achieved with the SBND proposed in \cite{bennatan} and the IED decoder using the DNN designed in \cite{kavv2}. It is observed that the result obtained in \cite{kavv2} already exceeds the performances shown in \cite{lugosch,beery}. In turn, with the same DNN and using the proposed IED decoder we achieve even better performance. For the interval $T\in\left[2,5\right]$, we observe a gradual improvement, reaching up to $0.7$ dB (for $T=5$) compared to the result obtained in \cite{kavv2}, when $\hbox{BLER}=10^{-3}$. Our tests indicate that, for $T>5$, the improvement is not significant. 

\subsection{BCH(63,36) code}
\label{ssec:experiments-36}
For the BCH(63,36) code, we propose the 8-layer architecture, with seven fully connected hidden layers, each of which has $8n=504$ units and uses the logistic sigmoid activation function. We also include a single skip connection (concatenation) from the first to the fourth layer. All hidden layers are followed by batch normalization layers to help with the stability and acceleration of the learning process \cite{fellow}.

For the learning rate, we obtained our best results by applying a triangular cyclic schedule\cite{smith} with minimum at~$10^{-5}$, maximum at $10^{-3}$, and a half-cycle of 64 iterations.

\begin{figure}[htbp!]
	\centering
	\begin{subfigure}[b]{0.48\textwidth}
		\centering
		\includegraphics[width=0.95\linewidth]{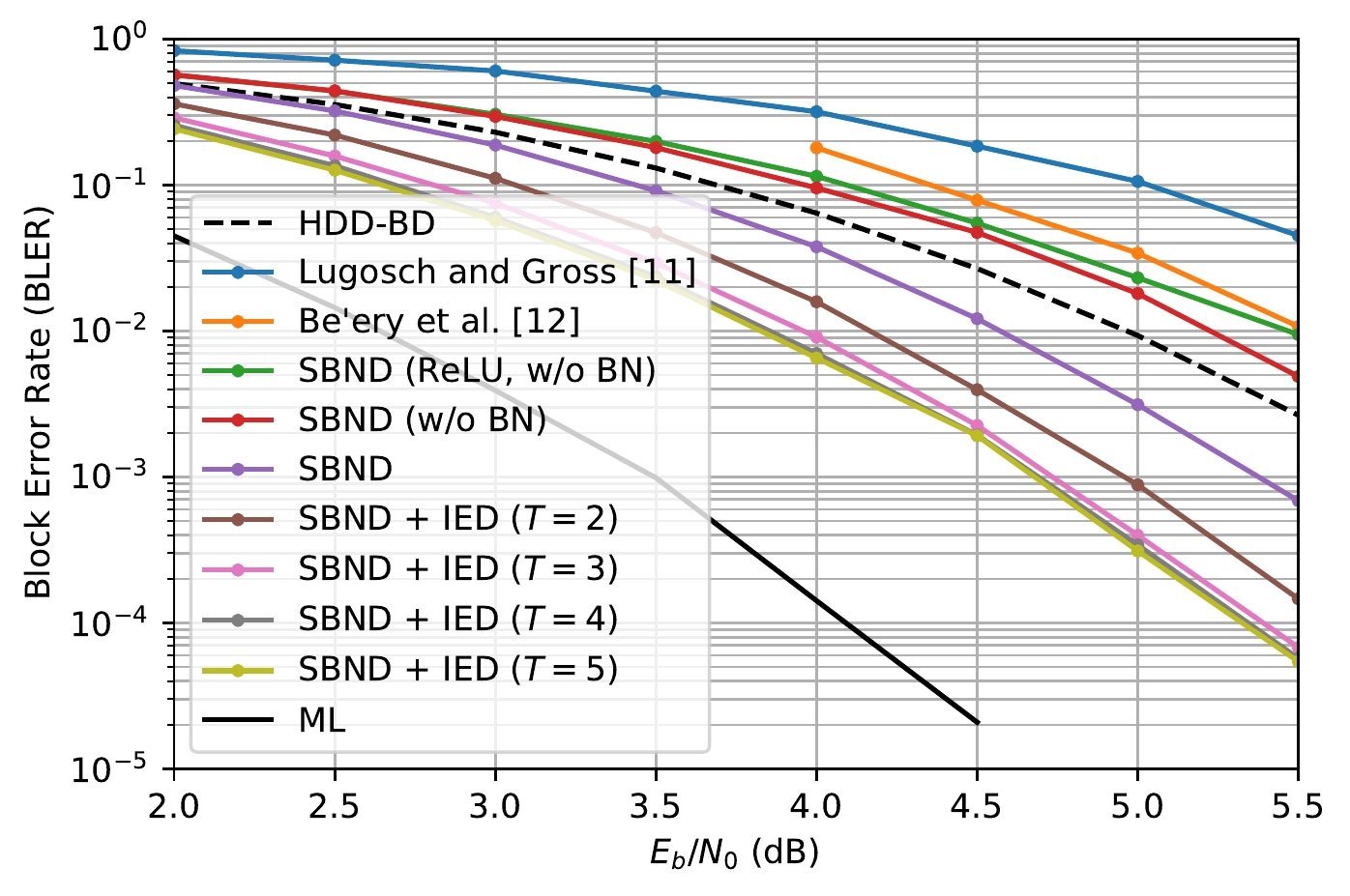}
		\centering
		\caption{}
		\label{fig:res_IIa}
	\end{subfigure}
	\hfill
	\begin{subfigure}[b]{0.48\textwidth}
		\centering
		\includegraphics[width=0.95\linewidth]{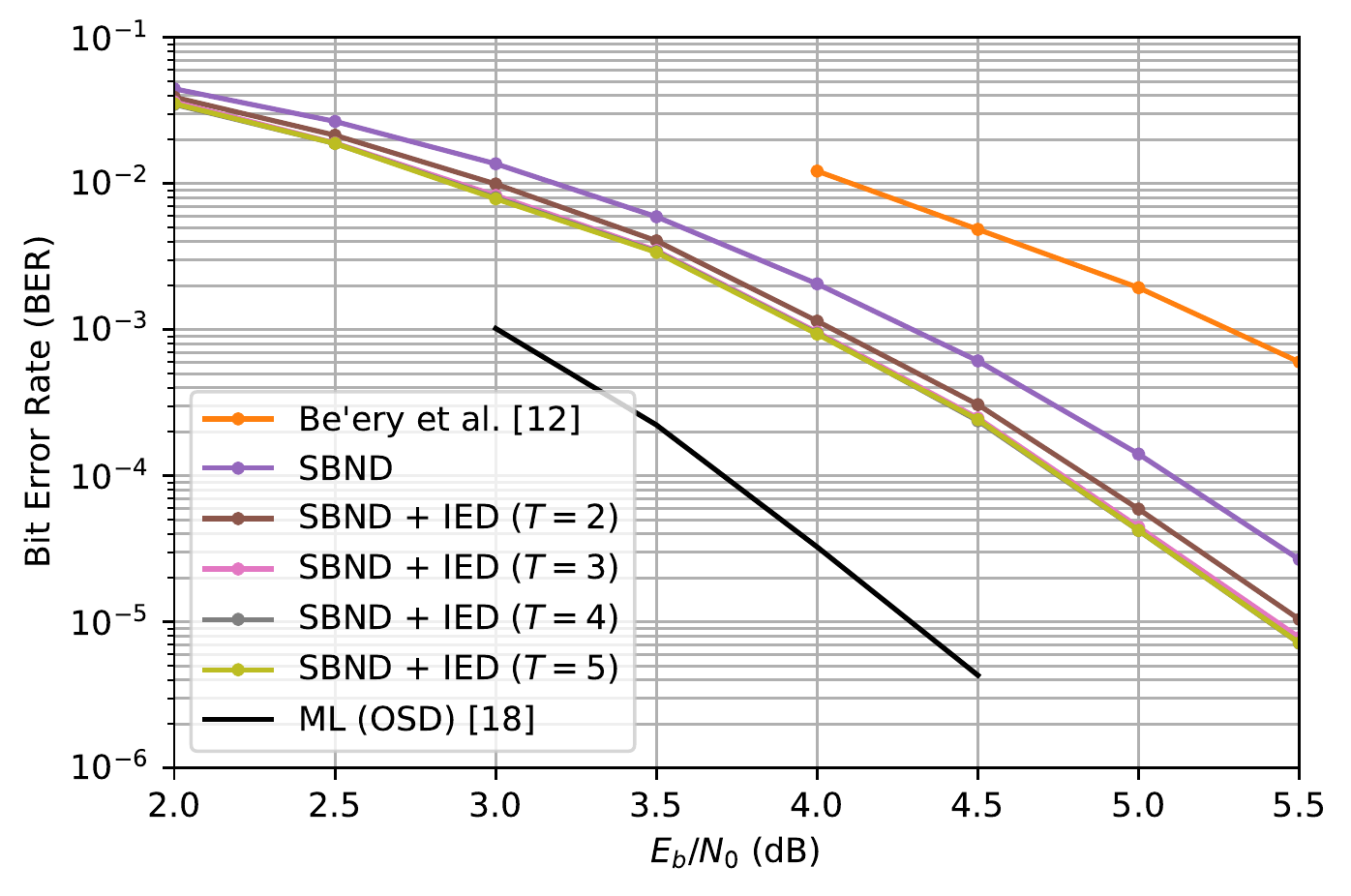}
		\centering
		\caption{}
		\label{fig:res_IIb}
	\end{subfigure}
	\caption{Performance obtained with the decoder in \cite{bennatan} and the IED decoder for the BCH(63,36) code using the DNN proposed in section \ref{ssec:experiments-36}.}
	\label{fig:res_II}
\end{figure}

Fig. \ref{fig:res_II} shows the performance of the proposed DNN with the decoder in \cite{bennatan} and the IED decoder. In Fig. \ref{fig:res_IIa}, ``(w/o BN)'' indicates a version with the batch normalization layers removed and ``(relu, w/o BN)'' indicates a further modification where the sigmoid activation of the hidden layers is replaced by ReLU. We can see that the combined use of the sigmoid activation and the batch normalization layers significantly improves the performance.


Again, it can be seen that the IED decoder achieves better performance than the results in the literature, including those of \cite{bennatan}. As in the case of the BCH(63,45) code, our best result is obtained when $T=5$, providing a gain of approximately $0.8$~dB at $\hbox{BLER}=10^{-3}$.

\subsection{Comparison with the Syndrome Loss}

To investigate whether the problem of disturbance discussed in Section~\ref{ssec:ied-motivation} could be solved by simply penalizing syndrome violations (without IED), we have trained the DNNs of sections \ref{ssec:experiments-45} and \ref{ssec:experiments-36} using the decoder of \cite{bennatan} and the hybrid loss function proposed in \cite{lugosch}, which incorporates a syndrome loss component besides the BCE loss. We experimented training from scratch and after pretrainng with the BCE loss. However, in both cases, the results were worse than using only the BCE loss and therefore were not included in the figures. This is not surprising since the syndrome loss was proposed in the context of belief-propagation decoding. Moreover, it ideally implies committing to a single rather than multiple superimposed error vectors, which may simply be too hard to learn under an inconsistent training set. In contrast, the BCE loss makes no such commitment, allowing the first iteration of the IED decoder to find and flip the single bit that is most likely to be in error.

\section{Conclusion}\label{sec:conclusion}
In this letter, we proposed a new decoder that uses the knowledge of the syndrome vector to feed a DNN designed to estimate the error pattern, where a stage of selecting the most confident positions to correspond to errors is used in order to improve estimation of the transmitted codeword.  In addition, we designed a new DNN for decoding the BCH(63,36) code.

The results obtained for the BCH(63,45) and BCH(63,36) codes show that the new decoding algorithm improves the performance of the SBND presented in \cite{bennatan}, at the price of a moderate increase in complexity. 
The IED decoder is flexible in the sense that it can be directly applied to any syndrome-based neural decoder without retraining.

\bibliographystyle{IEEEtran}
\bibliography{IEEEabrv,biblio}
\end{document}